\documentclass[11pt,twocolumn]{article}

\usepackage{amsmath} 	
\usepackage{amsthm} 	
\usepackage{amsfonts} 	
\usepackage{bm} 		
\usepackage{graphicx}
\usepackage[labelfont={it},textfont=it]{caption}
\usepackage{subcaption}

\usepackage{titlesec}
\titleformat*{\section}{\Large\bfseries}
\titleformat*{\subsection}{\Large }

\usepackage{url}

\usepackage{breakurl}
\usepackage[breaklinks]{hyperref}

\usepackage[margin=13mm]{geometry}

\usepackage{abstract} 

\renewenvironment{abstract}
 {\small
  \begin{center}
  \bfseries \abstractname\vspace{-.5em}\vspace{0pt}
  \end{center}
  \list{}{%
    \setlength{\leftmargin}{0mm}
    \setlength{\rightmargin}{\leftmargin}%
  }%
  \item\relax}
 {\endlist}

\usepackage{fancyhdr}
\usepackage{xcolor}
\usepackage{etoolbox}
\makeatletter
\patchcmd{\f@nch@foot}{\rlap}{\color{gray}\rlap}{}{}
\patchcmd{\footrule}{\hrule}{\color{gray}\hrule}{}{}
\makeatother

\title{ {\LARGE \textbf{Cyber C2: Achieving Scrutability and Agency in Cyberspace Operations}}}
\author{ Daniel Salmond, 
Van Nguyen,
Anton V. Uzunov,
Natalia Nikolova,
Prajakta Desai,
Ross Kyprianou \\ \emph{Defence Science and Technology Group} }
\date{\vspace{-5ex}}

\begin{document}
\twocolumn[
  \begin{@twocolumnfalse}
    \maketitle
    \thispagestyle{fancy}  
        \begin{abstract}
            {\small {\bfseries 
            
Our thesis is that operating in cyberspace is challenging because cyberspace exhibits extreme variety, high malleability, and extreme velocity. These properties make cyberspace largely inscrutable and limits one's agency in cyberspace, where agency is the ability to exert influence to transform the state or behaviour of the environment. With this thesis, we explore the nature of cyberspace, C2 and diagnosis of the challenges for cyber C2, with treatment to follow in future work. 

For the C2 expert unfamiliar with cyber, we consider definitions of cyberspace within military doctrine and the open literature. For the purposes of the paper, we adopt a definition that emphasises the digitised nature of cyberspace, admits human behaviours that influence or are influenced by cyberspace, includes logical electromagnetic interactions by radios, and includes the quality of being self-referent in that acting upon cyberspace can create and destroy cyberspace.

Similarly, for the cyber expert unfamiliar with C2, we explore the different interpretations of C2 within both doctrine and the research community. We note the increasing realisation that extant C2 approaches are not fit-for-purpose for increasingly complex operations, which are increasingly complex because of the interconnectedness afforded by cyberspace.

The unique challenges for the C2 of cyberspace are a consequence of the variety, malleability and velocity of cyberspace and lead to inscrutability and loss of agency. There is a clear need for structures, processes and technologies with the requisite variety, malleability and velocity to improve scrutability and maximise agency in cyberspace. Whilst international research efforts are underway to build better models and tools for doing so, we believe that by addressing variety, malleability and velocity explicitly, we lay a foundation for the development of control structures that maximises agency.  
            }
            }
        \end{abstract}
        \vspace{20pt}
  \end{@twocolumnfalse}
]

\section{Introduction}\label{sn:intro}
Rapid technological developments exploiting cyberspace and the electromagnetic spectrum (EMS) have brought forth both opportunities and threats, as sophisticated adversaries become more complex, destructive, and unpredictable \cite{black2020cyber}. Accordingly, cyberspace was formally recognized and declared as an operational domain by the US Department of Defense (DOD) in 2011 \cite{morgan2019command}, by the North Atlantic Treaty Organization (NATO) at the Warsaw Summit of July 2016 \cite{NATO2016}, and in the Defence Strategic Update 2020 \cite{DSU2020}. 

The logic for this speaks to the prominence of cyberspace to military operations. If one cedes control of cyberspace to an adversary, then one can no longer assure the availability, integrity or confidentiality of the information that flows therein. Every decision based on that information is at risk of denial and manipulation. By direct analogy, ceding control of the maritime domain means one can no longer assure the flow of materiel (oil, foodstuffs, chemicals for the manufacture of munitions) in a time of war. Ceding control of the space domain means you can no longer assure the provision of space-based services, such as satellite communications, or positioning, navigation and timing services. 

This has lead to a shift away from viewing cyberspace as merely an enabler of operations in other domains to acknowledging cyberspace as a domain in its own right, through which deterrence and coercion can be practiced and decisive kinetic and non-kinetic effects delivered \cite{black2020cyber}. As such, the intrinsic nature of warfare in cyberspace does not differ from warfare in other contexts \cite{quinter2012joint}. Consequently, it has been argued that the command and control (C2) of cyberspace needs to share common abstractions and frameworks with the C2 of other warfighting domains so as to support integrated operations~\cite{Routier14}. Accordingly, the 2023 Defence Strategic Review (DSR) recommends that~\cite[p64]{DSR2023}

\begin{quote}
\textit{[a] comprehensive framework should be developed for managing operations in the cyber domain that is consistent with the other domains}.
\end{quote}

On the other hand, it has been suggested that traditional C2 doctrine be re-examined, noting that cyberspace underpins much of the information and operational technologies required to support modern military operations. Existing C2 constructs and structures may need to be adapted to reflect the temporal, relational and spatial differences presented by cyberspace so as to enable the speed and agility required to keep pace with change in cyberspace. Other teams are considering how the C2 of cyberspace operations integrates with multi-domain operations~\cite{morgan2019command,Grant23,KaBo24}, however, in this paper we seek to understand the unique challenges for the C2 of cyberspace operations and so adopt a narrower scope. We will leave to the reader to extrapolate the consequences of our thesis to the integration of cyberspace operations with those in other domains.

The DSR asserts that~\cite[8.56]{DSR2023} 
\begin{quote}
\textit{Australia's cyber and information operations capabilities must be scaled up and optimised.}
\end{quote}



Given the nascent quality of cyber C2 research within Australia, this paper seeks to stimulate a dialogue with you, the reader, on the nature of cyberspace and the C2 of cyberspace operations. We proceed on the basis that this paper is conceptual framing and diagnosis of the challenges for cyber C2, with treatment to follow in future work. Our thesis is that operating in cyberspace is challenging because cyberspace exhibits extreme variety, high malleability, and extreme velocity. These properties make cyberspace largely inscrutable and limits one's agency in cyberspace, where agency is the ability to exert influence to transform the state or behaviour of the environment. 

Section 2 will introduce working definitions for cyber and cyberspace, based on extant doctrine and literature, and discuss its variety, malleability and velocity. Section 3 will present different interpretations of C2 and introduce the importance of scrutability and agency. Section 4 will explore the unique challenges for C2 of cyberspace operations. Section 5 will present final conclusions and advocate for the need for research that addresses the fundamental nature of cyberspace in order to increase scrutability and agency in cyberspace operations.

\section{Cyber and Cyberspace}\label{sn:cyber}

The term `cyber' is widely used both within the Australian Defence Force (ADF) and elsewhere. When pressed for a definition of cyber, cyber professionals offer a diverse range of definitions\footnote{To support the development of the Defence Cyber Science and Technology Strategy (2013)~\cite{defcyberstrategy}, the lead author interviewed over 60 members of industry, academia and government both within Australia and overseas both on the nature of cyberspace and the science and technology challenges for cyber research in Defence.}. 

As Moore~\cite{Moore2019} notes, 
\begin{quote}
\textit{[i]n its most abstract, appending cyber as a prefix simply means “involving a 
computer”. A reasonable concern is that as most human functions and interactions become more 
reliant on some form of computed involvement, the term itself becomes redundant.}
\end{quote}

Many publications eschew giving a definition of cyber, or adopt definitions that are complicated and unwieldy. Comparison of definitions between communities reveal that there is a diversity of opinion on what is and is not regarded as cyber.  

We believe that cyber is not well-defined because of three reasons: a) cyber is a relatively new concept, and the mechanisms of language convergence around concepts can take decades if not centuries to play-out for concepts that lack tangibility; b) the concept of cyber is closely related to the rise and prevalence of information systems, which themselves are undergoing constant change; and c) most abstract concepts are best described in terms of less abstract concepts, which themselves are described in less abstract concepts. We posit that the hierarchy of conceptual abstraction is itself lacking and evolving. Without a clearer definition, we risk adopting the same approach as for the definitions of `life' and `machine intelligence', i.e. ``you'll know it when you see it''.

Consequently, we present a number of definitions and interpretations of cyber from doctrine and the literature. Our goal is to identify a working definition for the purpose of exploring the C2 of cyberspace operations in the remainder of this paper.

The United Stated Joint Publication on Cyberspace Operations~\cite{jp-3-12} defines cyberspace as

\begin{quote}
\textit{A global domain within the information environment consisting of the interdependent network of information technology infrastructures, including the Internet, telecommunications networks, computer systems, and embedded processors and controllers}
\end{quote}



The former Australian Head of Information Warfare Division, MAJGEN Coyle, defined cyberspace as cyberspace~\cite{Coyle21}:
\begin{quote}
    \textit{the global digital environment of partitioned and interdependent logical and hardware infrastructure, networks, systems, information and services}
\end{quote}
and the ADF Capstone Doctrine expressed in `Australian Military Power'~\cite{ADFC0} asserts that 
\begin{quote}
\textit{Cyberspace consists of all interconnected communication, information
technology and other electronic systems, networks and their data,
including those which are separated or independent.}
\end{quote}

The common theme across these two ADF definitions is that cyberspace is largely a logical construct made up of physical devices and networks, and logical abstractions provided by protocols and services.

The United Stated Joint Publication on Cyberspace Operations~\cite{jp-3-12} identifies cyberspace as having three distinct, inter-related layers. The physical layer consists of the devices and infrastructure that constitute the physical manifestation of cyberspace. The logical network layer consists of the logical abstractions that dictate the specified and emergent behaviour of cyberspace. Finally, the cyber-persona layer consists of the web of user and system accounts via which cyberspace is accessed and manipulated. We note that this three layer model places the human users outside the scope of cyberspace.

\begin{figure}
\centering
\includegraphics[width=9cm]{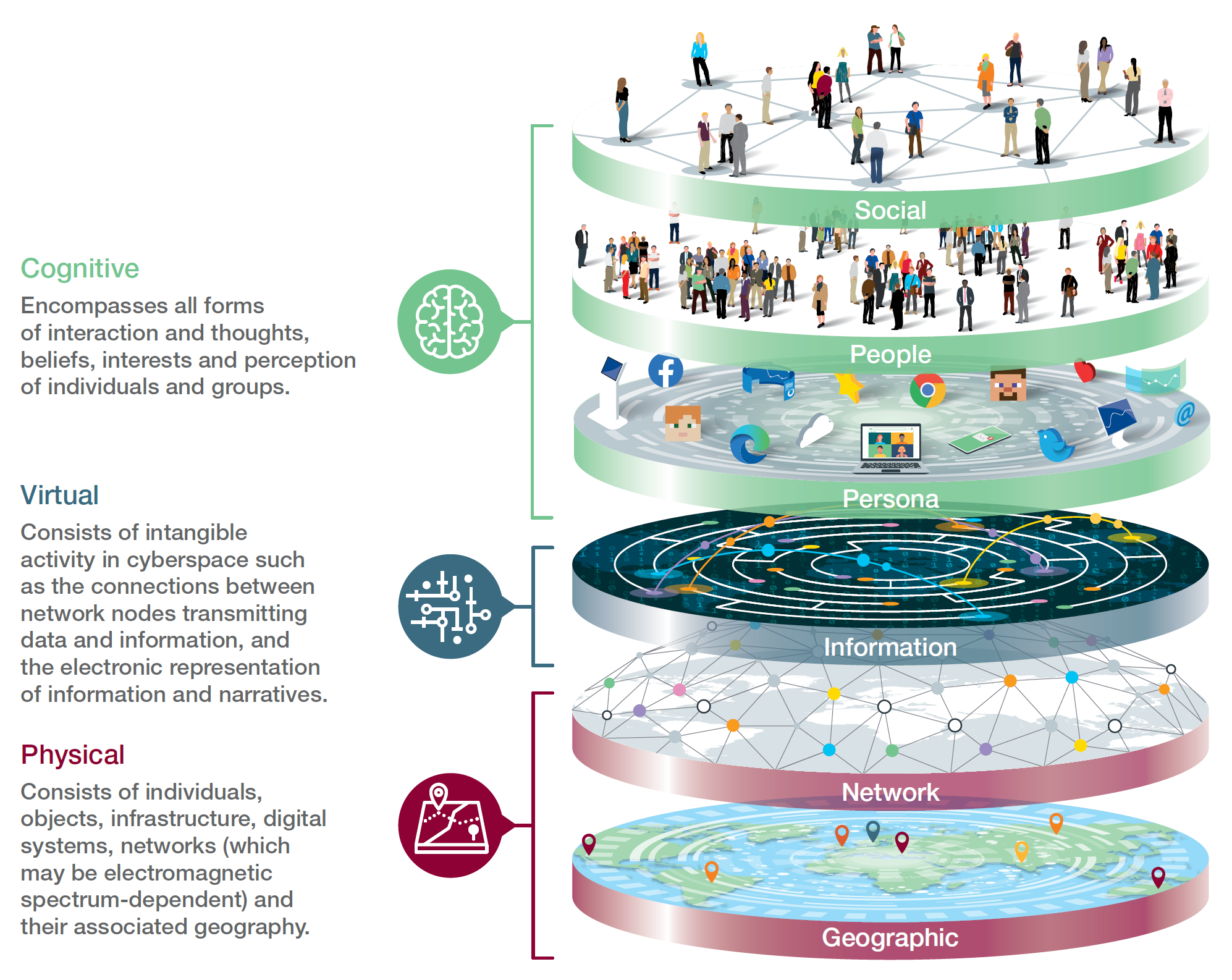}
\caption{The UK Cyber Primer's six layers of cyberspace~\cite{cyberprimer}}
\label{fg:cyberlayers}
\end{figure}

In their Cyber Primer~\cite{cyberprimer} the United Kingdom (UK) Ministry of Defence (MOD) adopts a similar three-dimensional model as the United Stated Joint Publication on Cyberspace Operations. Their model, illustrated in Figure~\ref{fg:cyberlayers}, consists of physical, virtual and cognitive dimensions, divided into a total of six layers. Relative to the US model, the UK model splits the physical dimension into geographic and network layers, recognising that electromagnetic emissions propagate through geographic space in addition to the physical manifestation of devices. Their virtual dimension contains a single information layer analogous to the US model's virtual layer. The cognitive dimension includes personas, people and social layers, where the last two layers are outside of scope of the US model.

Further, the Cyber Primer defines cyber in terms of an Oxford English Dictionary definition,
\begin{quote}
\textit{relating to information technology, the Internet and virtual reality},
\end{quote}
which is evocative of Moore's thesis~\cite{Moore2019} that cyber pertains to computers and their networks. The Primer defines cyberspace as 
\begin{quote}
\textit{the global environment consisting of all interconnected communication, information technology and other electronic systems, networks and their data, including those which are separated or independent, which process, store or transmit data}
\end{quote}

The ADF and UK MOD definitions are similar despite their choice of wording. By contrast the US definition does not explicitly include the data or the processes associated with storage and processing of data. Despite the omission, we do not believe this is an oversight; it is difficult to conceive of cyberspace without the inherent data and processes that occur within it and so we assume that data and processes are implicit in the US definition.

The NATO Allied Joint Doctrine for Information Operations~\cite{nato_AJDIO} adopts a similar approach to the US by identifying three interdependent `effect dimensions' for information operations: cognitive, physical and virtual. Every military effect will have consequences in one or more of these dimensions. The virtual dimension refers to consequences that concerns `... the storage, content and transmission of analogue and digital data. It also includes all supporting communication and information systems and processes.' It is noteworthy that analogue data is included in the virtual dimension. The doctrine implies that cyberspace operations seek to act upon the digital aspect of the virtual dimension, with ramifications for both the cognitive and physical dimensions. 

The NATO doctrine further notes that cyberspace is constantly evolving, can be used by anyone for almost any purpose, and is entirely human-made. This leads us to conclude that, unlike other domains of warfare, cyberspace operations do not just occur within cyberspace but act on the substance of cyberspace itself. The creation of new information capabilities is the creation of more cyberspace. Conversely, degradation of one's own or an adversary's information capabilities is a concordant degradation of cyberspace.

Clark~\cite{Clark10} defines cyberspace\footnote{Clark notes that the term `cyberspace' was coined by William Gibson. Gibson's first use of the term is in a 1982 \textit{Omni} magazine story then in his 1984 novel, Neuromancer.} as being comprised of people (users, creators, transformers of cyberspace), information (stored, transmitted, and transformed), logical structure, and physical embodiment. Clark asserts that the purpose of cyberspace is to support 

\begin{quote}
\textit{the processing, manipulation and exploitation of information, the facilitation and augmentation of communication among people, and the interaction of people and information}
\end{quote}

Relative to ADF, UK and US doctrine, Clark explicitly includes the users in their conceptualisation.

Clark further writes:

\begin{quote}\textit{The nature of cyberspace is the continuous and rapid evolution of new capabilities and services, based on the creation and combination of new logical constructs, all running on top of the physical foundations. Cyberspace, at the logical level, is thus a series of platforms, on each of which new capabilities are constructed, which in turn become a platform for the next innovation. Cyberspace is very plastic, and it can be described as recursive; platforms upon platforms upon platforms. The platforms may differ in detail, but they share the common feature that they are the foundation for the next platform above them.}
\end{quote}

We note the strong correlation between Clark's view of cyberspace and Salmond et al.'s following definition of information warfare~\cite{SaTrNiIg23}: 

\begin{quote}\textit{the manipulation of information flows and information processing systems (human or machine) in order to influence human and machine-decision making, while preserving and enhancing the integrity and availability of one's own decision making}
\end{quote}

In this vision of information warfare, the information environment is decomposed into elements, each of which conveys - and potentially transforms - information through space and time. 

While Clark's and Salmond et al.'s definitions include people, the National Military Strategy for Cyberspace Operations~\cite{nmsoco} defines cyberspace as 

\begin{quote}
\textit{a domain
characterized by the use of electronics and the electromagnetic spectrum}
\end{quote}
and therefore includes the electromagnetic spectrum but not people. 

Whilst cognitive warfare aspects are excluded, it is well recognised that cognitive warfare effects may be delivered via cyberspace operations. This has lead Ducheine et al.~\cite{DuPiAr22} to distinguish between \textit{hard} cyberspace operations, which occur \textit{in} cyberspace, and \textit{soft} cyberspace operations, which occur \textit{through} cyberspace.

With electronic warfare (EW) operations included within cyberspace operations, one may ask whether there is a difference between purely cyber effects vs EW effects. An occasionally offered view in Australian and US Defence communities\footnote{To the best of the authors' knowledge, this view does not appear to be codified in any publication.} is that a pure cyber effect is one that leads to a persistent change in the targeted information system, whereas an EW effect is one that lasts only as long as the effector is active. Proponents of this view argue that jamming and spoofing signals are only effective as long as the jamming or spoofing system irradiates power than can be received by the target. When the irradiated power ceases, then the target receiver can (typically) resume normal operation. By contrast a cyber effect, such as flipping a bit, will persist after the cyber effect ceases. We reject this view: distributed denial of service attacks render a targeted web service unavailable for the duration of the attack, but the effect does not persist beyond the end of the attack. Moreover, the distinction between cyber and EW effects becomes blurred when electromagnetic effects can lead to persistent changes in the targeted information systems~\footnote{One reviewer noted that the burning out of an electromagnetic sensor would constitute a persistent electromagnetic effect.}. 

Routier~\cite{Routier14} reframes Clark's 4 layer model~\cite{Clark10} in terms of logical, physical, objective and actor layers. The emphasis on objectives and actors is striking but reasonable given that Routier was focused on C2 of cyberspace wherein objectives are first-order objects. The broader term `actors' is also especially apt. In the intervening decade between Clark's and Routier's publications, machine-based processes were increasingly and autonomously creating and transforming of cyberspace. Routier advocates that this 4 layer model should be adopted and applied across multi-domain C2.

Routier also frames cyberspace in terms of the Open Standards Interface (OSI) stack~\cite{osi}, which defines a seven layer model for the mediation of interactions between users, applications and physical devices. It, or rather the protocols within an OSI stack, are responsible for a user being able to speedily and reliably access content on the internet without knowing where that content is physically stored. Routier recommends that cyber C2 systems should ignore the physical layer of the OSI stack, but should provide complete situational awareness for all six of the remaining layers. 

Thus, we adopt two working definitions for the remainder of this report. We follow Moore's lead in defining `cyber' as a prefix
indicating that object relates to computers and computer networks.  We shall define `cyberspace' as  
\begin{quote}\textit{the interacting and interdependent union of computational devices, logical protocol stacks, and emergent system behaviours -- whether machine-centric or at the human-machine interface -- that support the generation, storage, transmission, and processing of digitised information, protocols and emergent system behaviours.}
\end{quote}

This definition emphasises the digitised nature of cyberspace, admits human behaviours that influence or are influenced by cyberspace\footnote{... but otherwise excludes general human behaviour.}, includes logical electromagnetic interactions by radios, and includes the quality of being self-referent in that acting upon cyberspace can create and destroy cyberspace. From here on, we shall refer to cyberspace operations in lieu of cyber operations to underscore that such operations have broader scope than just computer networks. The term cyber C2 will be used as shorthand for the C2 of cyberspace operations.

We posit that there are three fundamental and inter-related properties that make cyberspace unique relative to other operating environments\footnote{In previous work~\cite{SaTrNiGr23}, Salmond et al. advanced that the information environment also exhibited other properties, e.g. non-locality and non-stationarity. However, we now believe that the  properties of variety, malleability and velocity are the fundamentals from which other properties can be derived. The perception that cyberspace affords non-local action is actually a consequence of the variety and velocity of cyberspace. Similarly, non-stationarity is a consequence of the variety and velocity of cybersapce.}: \textit{variety}, \textit{malleability} and \textit{velocity}.

We refer to \textit{variety} in the sense of Variety Calculus~\cite{NiCa23}, which is inspired by Ashby's notion of variety in cybernetics~\cite{Ashby57}\footnote{We observe that the Variety Calculus is a qualitative, systems theory framework. By contrast, Ashby's definition is framed in reductionist, quantitative terms.}. Niven and Capewell describe variety as~\cite{NiCa22}
\begin{quote}
    \textit{... a characteristic of a system derived from the diversity of components and interactions of which it is composed. Greater Variety means that a system is more complex, more disordered, and requires more information\footnote{Explicitly, one could define variety as the amount of information required to describe a system, including both the possible states of its components and the behaviour that regulates interactions between them. We note that our model of influence~\cite{Salmond23,Salmond23b} explicitly represents this information for a given system and is thus well-disposed to report the variety of the given system. } to enable understanding.}
\end{quote}
The system is regarded as complex once the variety of the system exceeds one's ability to comprehend the total state and operation of that system~\cite{NiCa23}. Cyberspace exhibits extreme variety relative to other domains because of the extreme interdependence. Moreover, the variety of cyberspace is effectively boundless: the manifestation of cyberspace and their behaviours is limited only by the technology stack of the day. 

This is not to suggest that other domains, e.g. society, are not boundless in terms of their potential variety. In fact, the increasing complexity of modern society has been attributed to the emergence of cyberspace~\cite{NiCa22}:
\begin{quote}
\textit{the ubiquity of sophisticated communications and information technology has established greater connectivity across the scope of human activities. Complexity arises through such connections, influences and dependencies.... Seemingly unconnected or geographically remote events can have profound influences across these networks of relationships...}    
\end{quote}
The variety of cyberspace can be traced to the foundational contributions of Church~\cite{Church36} and Turing~\cite{Turing36} to computer science. Among other things, they showed that finite means can produce infinite scope (variety!): a discrete alphabet, whether binary, ASCII, etc., can be composed and recomposed to produce behaviours of unbounded and undecidable complexity\footnote{In fact, the non-stationary and self-referential nature of cyberspace is a direct analogue to the setting of Turing's halting problem~\cite{Turing36}.}. Unlike other domains, the manifestation of that infinite scope is amenable to construction and analysis; it can be examined and brought about more easily than in other human endeavours.

The \textit{malleability} of cyberspace, i.e. the ease with which it can be transformed, is also unique. Under the right circumstances, a carefully chosen set of bit-flips can render a system inoperable or redirect a significant proportion of internet traffic: small changes can have dramatic consequences~\cite{AnAn22}. Similarly, software developers can publish patches that eliminate vulnerabilities. New technology stacks can transform the way users interact with each other, with systems and with data. This is not to suggest that finding vulnerabilities to exploit or to patch is easy; the variety of cyberspace negates this. However, the malleability of cyberspace means that it can shift dramatically, exacerbating the variety of cyberspace.

Finally, the \textit{velocity} of cyberspace, i.e. the speed at which information both propagates and is processed, is the basis for our hyper-connected information age. The speed at which these interactions can occur means that human interaction is increasingly mediated by more cyberspace elements such as content curation algorithms, web-store recommendation systems, predictive caching of anticipated search results, and code autocompletion routines. Interactions with cyberspace itself will increasingly require yet more sophisticated tools (yet more cyberspace) to facilitate those interactions, taking advantage of the ever-evolving useful interfaces while fending off the exploitative ones.

\section{Command and Control}\label{sn:c2}
Command and control (C2), like cyber, is  subject to many interpretations. Our purpose in this section is to provide the reader with a set of interlocking interpretations of C2 so that we can discuss the unique challenges for C2 of cyberspace operations in Section~\ref{sn:cyberc2}.

We firstly note that C2 has a specific meaning in cybersecurity. It refers to the set of techniques by which an adversary communicates with compromised systems under their control~\cite{mitrec2}. The notion of C2 discussed in this paper is significantly broader.

Vassiliou, Alberts and Agre regard C2 as~\cite{VaAlAg15}
\begin{quote}\textit{the set of organizational and technical attributes and processes by which an enterprise marshals and employs human, physical, and information resources to solve problems and accomplish missions.}
\end{quote}
Although the term `C2' is typically used in military parlance, it should be better understood as four distinct functions. These are:
\begin{itemize}
    \item Command: the expression of intent, and the exercise of authority and the delegation thereof in pursuit of operational objectives 
    \item Control: the execution, monitoring and correction of operations, with respect to the commander's intent
    \item Coordination: planning, synchronisation and deconfliction of operational activities, including their sequencing, timing and tempo
    \item Communication: the passing of information between operational elements
\end{itemize}
The latter two functions are typically implied by the former. As such, extant doctrine focuses on the terms `command' and `control' in their pre-eminent C2 documents.

NATO defines~\cite{natoglossary} command as
\begin{quote}\textit{the authority vested in a member of the armed forces for the direction, coordination, and control of military forces}\end{quote}
and control as
\begin{quote}
\textit{the authority exercised by a commander over part of the activities of subordinate organizations, or other organizations not normally under [their] command, that encompasses the responsibility for implementing orders or directives}\end{quote}.

Australian doctrine~\cite{addpc2} adopts a similar definition for command:
\begin{quote}\textit{the authority which a commander in the military Service lawfully exercises over subordinates by virtue of rank or assignment}\end{quote}
wherein the term `lawfully' refers to the constitutional and legislative accountabilities associated with command. The ADF and NATO definitions for control are essentially identical and focus on the role of authority in the exercise of C2. Alternative interpretations of C2 focus on control as direction, execution, monitoring and correction of a course of action, and cleaves more closely to the concept of control as found in, say, control theory.

Australian doctrine~\cite{addpc2} also notes that C2 occurs within operational environments that are typically complex, and differentiates between structural and interactive complexity. Structural complexity exists in a system with many parts. In the absence of interactions between these parts, the system is typically regarded as predictable. By contrast, interactive complexity occurs in systems with interactions between the parts, which may lead to non-stationary and unpredictable behaviour. 

Military affairs are increasingly complex because of the variety of elements and associated relationships in modern military environments, including the prevalence of coalition operations, the increasingly integrated nature of military and civilian affairs, and the emphasis on sub-threshold warfare~\cite{NiCa23}. Significantly, cyberspace underpins all three causes: the structural and interactive complexity of cyberspace imbues military operations with commensurate complexity.

The principal consequence of this increased complexity is that traditional, i.e. rigid and hierarchical, C2 structures are no-longer fit-for-purpose. Niven~\cite{Niven23} advocates for
\begin{quote}
\textit{... a shift from directive control of a highly structured force towards maintaining the purposefulness of more independent and diverse networks of actions and actors... The purpose of C2 is to ensure the purposefulness, coherence and effectiveness of collective action within an operating enterprise through the design, maintenance and regulation of operations.}
\end{quote}

This interpretation de-emphasises authority in favour of coherence of collective action; purposefulness becomes the driver of coherent action in lieu of authority. 

Whilst this deviates from the doctrinal conceptualisations of C2, the international C2 research community has been advocating for more flexible C2 approaches for the last two decades. In fact, this community has preferred to eschew definitions of C2 and opt instead for a characterisation of approaches that illuminate the C2 possibility space. For example, Alberts and Hayes~\cite{alberts2006understanding} argue that the C2 approach is contingent on the following characteristics of the operational environment:
 \begin{itemize}
\item{\textit{Rate of Change}}: refers to the extent of dynamism or stability within a situation. Static scenarios lend themselves to centralized decision-making, optimizing preplanned efforts under tight control. In dynamic circumstances, rapid changes render conventional controls impediments to effective command and control. 
\item {\textit{Degree of Familiarity}}: reflects the depth of comprehension surrounding a problem. High familiarity indicates a well-understood issue. However, this does not imply that dynamic situations are inherently less understood. 
\item {\textit{Strength of Information Position}}: denotes an organisation's capacity to fulfill its information needs. 
\end{itemize}
 
\begin{figure}
\centering
\includegraphics[width=9cm]{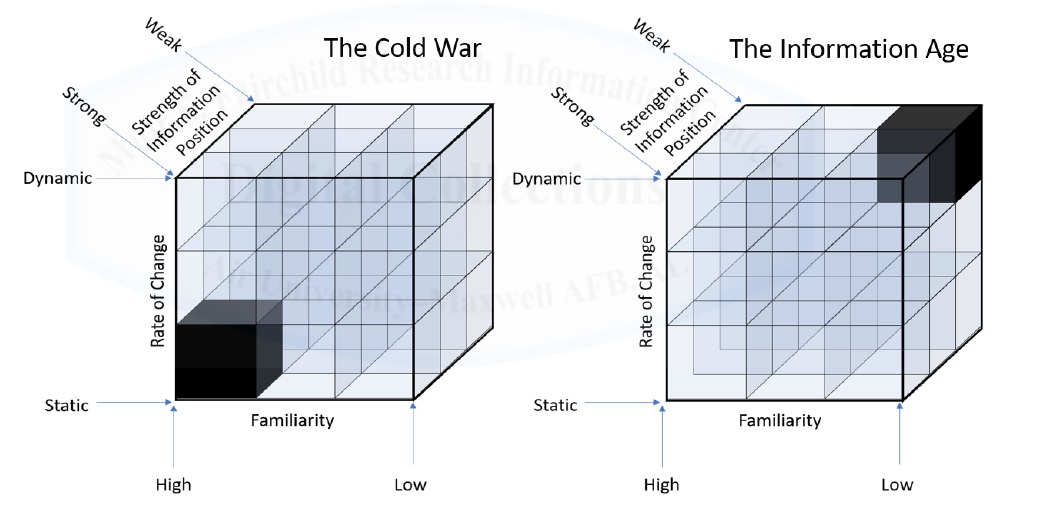}
\caption{The Spectrum of Operational Environments (from~\cite{alberts2006understanding})}
\label{fig:OE}
\end{figure}

Based on the characterization provided above, the nature of operational environments in the information age relative to the Cold War are illustrated in Figure \ref{fig:OE}.  The operational environment in the Cold War era is static with a high degree of understanding of both friendly and adversary capabilities. Conversely, information age warfare is conducted within a volatile, uncertain, complex, and ambiguous context, often against an unfamiliar or unknown adversary. Therefore, the operational environments in the Information Age are dynamic, featuring a lower degree of familiarity with the adversary and a relatively modest capacity to meet information requirements \cite{french2018operational}. 

Based on their model for operational environments, Alberts and Hayes  \cite{alberts2006understanding} identify the critical dimensions of a C2 framework: \textit{decision rights allocation}: describing how the command function will operate within the framework, \textit{interaction patterns}: involving factors such as the quantity and diversity of participants, the caliber of interaction content, and the methods utilized to facilitate the interaction, and \textit{information distribution}: describing how collaboration is enabled through the access to information. 

\begin{figure}
\centering
\includegraphics[width=9cm]{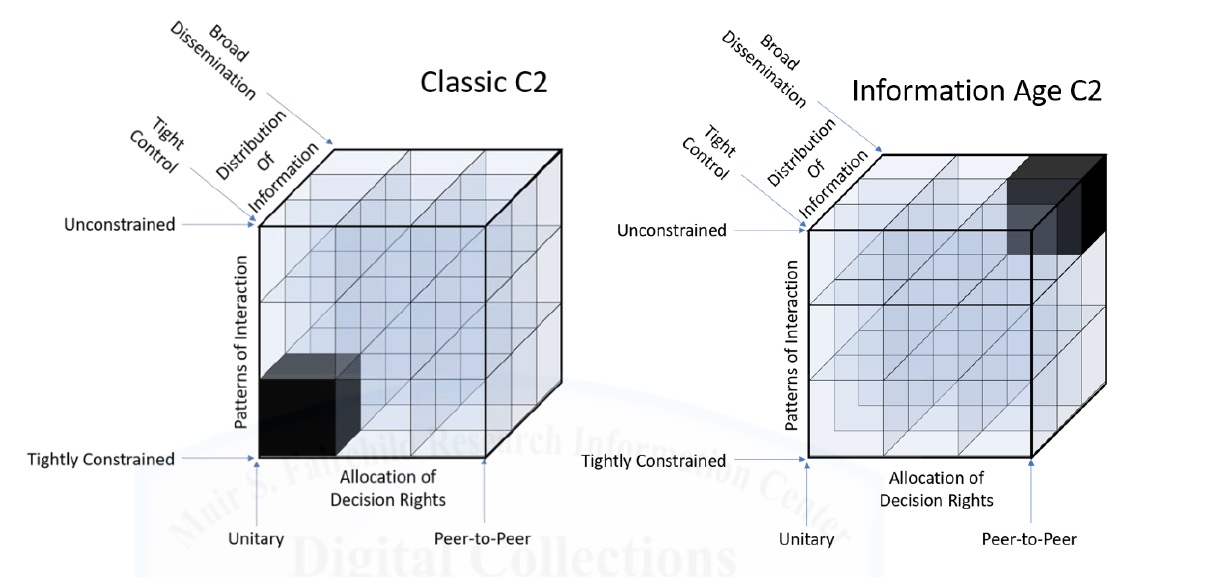}
\caption{The Spectrum of C2 Frameworks (from~\cite{alberts2006understanding})}
\label{fig:C2Framework}
\end{figure}

Figure \ref{fig:C2Framework} illustrates the contrast between classic and information age C2 frameworks. Classic C2 follows rigid hierarchical structures, with top-down decision-making and tightly controlling information flow. Information Age C2, on the other hand, promotes flattened organizational setups, enabling widespread vertical and horizontal collaboration, easy access to information, and loose controls for information sharing.  

The push towards less rigid C2 approaches has been under the moniker of \textit{agile C2}. The Australian response to the call for agile C2 has been the development of the ADF Concept for C2 of the Future Force~\cite{hcac}. The central premise of this concept paper is that the ADF should adopt a hierarchical command, agile control approach. The paper asserts that mission command~\cite{addpc2}, 
\begin{quote}
\textit{[a] philosophy for command and a system for conducting operations in which subordinates are given clear direction by a superior of their intentions... The result required, the task, the resources and any constraints are clearly enunciated, however subordinates are allowed the freedom to decide how to achieve the required result,    
}\end{quote}
remains a viable and hierarchical approach to command in the ADF. Agile control will be realised through lateral and collaborative pathways in contrast to hierarchical pathways. Moreover, the paper posits that future conflict will be technological, subject to false information and contested information environments. Necessarily, control may be exercised by human or machine. Significantly, C2 processes themselves may constitute cyberspace and be vulnerable to cyber attack, and worthy of cyber defences.

Lambert characterises the ability for C2 systems to achieve coherent action in terms of awareness, intentionality, and capability\footnote{We note that this characterisation is consistent with ADF doctrine~\cite{addpc2}. Lambert's thesis was that C2 functions could be performed by machines as well as humans, which we believe was contentious at the time.}~\cite{Lambert99}. Battlespace awareness is essential for planning and responding to exigencies. The ability to articulate commander's intent is essential to establish clarity of purpose. However, awareness and intentionality are insufficient: a commander must have capabilities at their disposal in order to act upon the environment to fulfil their mission. Salmond~\cite{Salmond22} introduces a causal model and information theoretic measures to formalise the interactions between awareness, intentionality and the ability to influence the environment, and demonstrated that the environment must be \textit{scrutible} for reasoning about achieving one's intentions in that environment. 

In Variety Calculus terms~\cite{NiCa21}, an organisation must achieve \textit{requisite variety} with the environment in order for it become scrutible. An environment is inscrutible to an observer when that observer lacks the mental models, resources, structures and processes to make sense of the environment. We note that relative to a C2 organisation,  the environment includes both those objects under its command, as well as the external environment at large. Requisite variety can be achieved by either \textit{amplifying} one's own variety, i.e. acquiring the suitable mental models, resources, structures and processes, or by \textit{attenuating} the variety of the environment itself. Accordingly, military C2 processes and structures have amplified over time to deal with the increasing complexity of warfighting systems and the environment at large.

Variety attenuation occurs by seeking to impose structures and processes on the environment to increase its predictability. For example, training attenuates the variety of a group of individuals to form a corps of soldiers, who can then be directed by a single commander to achieve coherent action. Variety attenuation can also be achieved through focusing the scope of an operation, or by influencing the environment at large.  The term \textit{agency} in lieu of control can be adopted to reflect that one's degree of influence over the environment is contingent and not guaranteed~\cite{NiCa21}. Again, an organisation must achieve requisite variety in order to maximise its agency in the environment. It follows that achieving requisite variety is necessary for both scrutability and agency, and increasing scrutability is causally linked to maximising agency~\cite{Salmond22}. 

We conclude this section with an interpretation of command, control and C2 that elegantly unifies the doctrinal and agile interpretations~\cite{PiMc02}:
\begin{quote}
    \textit{Command is the creative expression of human will necessary to accomplish the mission; control is the structures and process devised by command to enable it to manage risk. C2 is the establishment of common intent to achieve coordinated action.}
\end{quote}

\section{C2 for Cyberspace Operations}\label{sn:cyberc2}

Having explored the definitions of cyberspace and C2 in the previous sections, we may now consider their synthesis. Our intent is to identify the unique challenges that differentiate the C2 of cyberspace operations from other domains, and how these relate to scrutability and agency. 

As noted by Scherrer and Grund \cite{scherrer2009cyberspace}: 
\begin{quote}\textit{Although it stands to reason that cyberspace operations share similar C2 elements as other warfighting domains such as organization; technical systems; and tactics, techniques, and procedures; we assert that the most effective C2 method for cyberspace operations will be heavily influenced by the nature of the domain itself and the environment within which it exists.}
\end{quote}

We underscore Scherrer and Grund's first point by noting that C2 of cyberspace operations shares much in common with other domains. The goal of cyberspace operations is to achieve human-centric outcomes, typically the preservation or improvement of welfare within a polity. The C2 of said operations requires coordinated action, subject to uncertainty of information and outcome, and therefore risk management. Commanders of cyberspace operations are accountable: they must comply with the normative values of the polity performing the operation, both formal and informal. In Western liberal democracies, they are beholden to government legislation, policy and regulation. The exercise of C2 of cyberspace requires understanding of the context, the assignment and deployment of capabilities, and the intention on behalf of the commander~\cite{Lambert99}. However, and critically for cyberspace, C2 is limited by the capacity of human individuals, and teams of individuals to understand, plan and execute missions. 

\begin{figure}
\centering
\includegraphics[width=9cm]{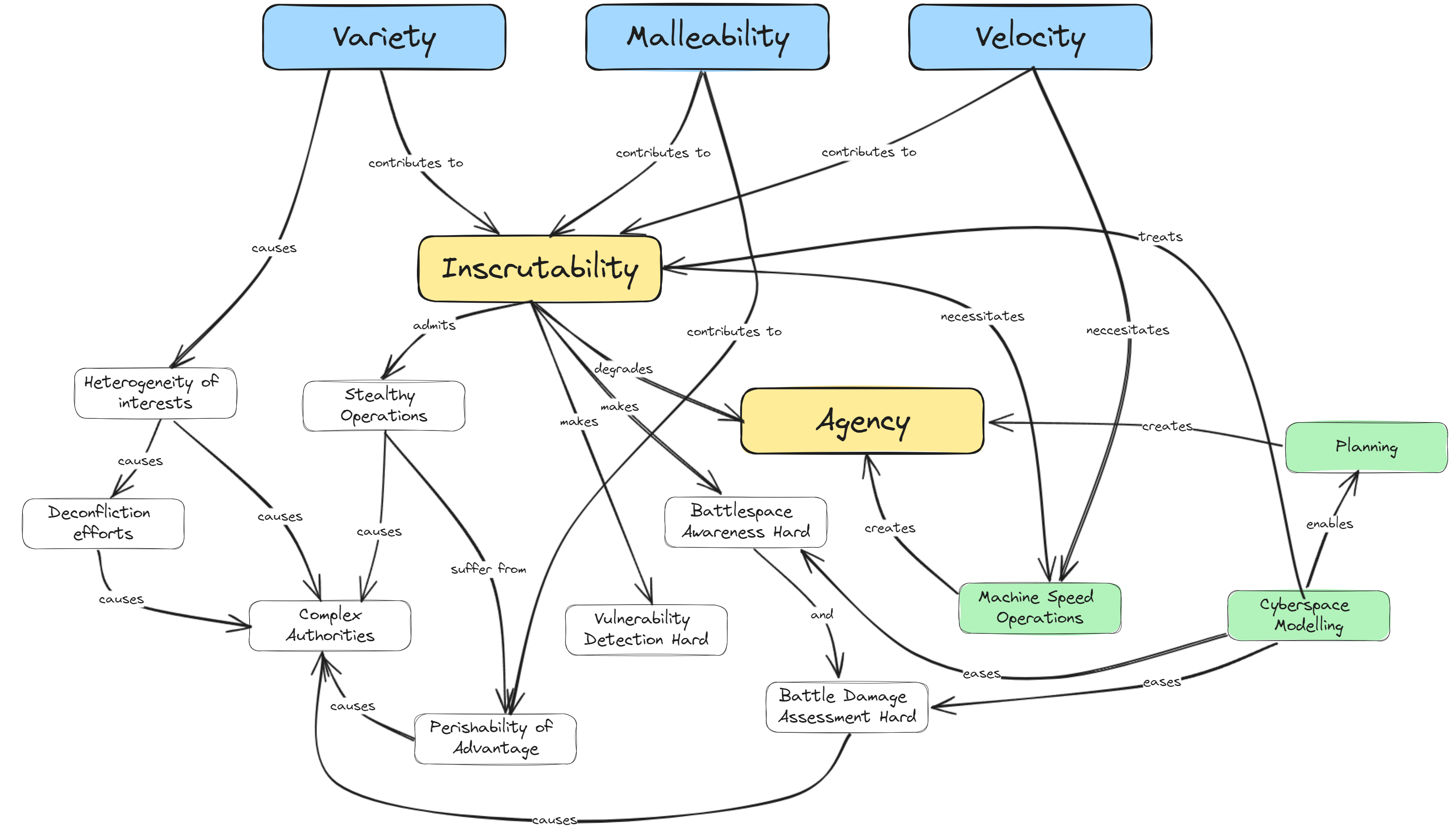}
\caption{Variety, malleability and velocity induce challenges for the C2 of cyberpsace operations.}
\label{fg:scrut-ag}
\end{figure}

We are not the first to identify the unique challenges for the C2 of cyberspace operations. For example, see~\cite{Grant23,Moore2019,morgan2019command,Routier14,scherrer2009cyberspace}. In fact, some would suggest that the nature of cyberspace is inimical to the application of traditional military strategy~\cite{KaCo17}. However, we argue that these occur as a consequence of the three properties of cyberspace identified in Section~\ref{sn:cyber}: variety, malleability, and velocity. We describe how these properties induce issues for scrutability and agency in cyberspace in the remainder of this section, as illustrated in Figure~\ref{fg:scrut-ag}. We posit the need for planning, cyberspace modelling and machine speed operations to address these challenges.

The variety of cyberspace may be partly attributed to the high levels of its interdependence\footnote{We note that this interdependence is by design: it is a feature, not a bug.}. Interdependence leads to heterogeneity of interests~\cite{sorensen2010cyber}; the actions of one entity in cyberspace can have far-reaching and unintended consequences. Consequently, coherent cyberspace operations necessitate coordination and de-confliction~\cite{Routier14}, e.g. between and within government agencies charged with monitoring or operating within cyberspace. 

The variety and velocity of cyberspace contribute to its \textit{inscrutability}. Cyberspace is complex because of the extreme variety of the objects and relationships that make up cyberspace, and the speed at which cyberspace operates. Human operators lack the cognitive bandwidth necessary to comprehend the state of cyberspace. Although machines can augment human cognitive bandwidth, they too lack the communications and processing bandwidth to digest and summarise cyberspace outside of a very limited scope. 

Inscrutability leads to poor situational awareness in cyberspace operations, which in turn affords the ability for stealthy operations~\cite{sorensen2010cyber}, whether criminal or state-based. Battle damage assessment, i.e. evaluating the effectiveness of cyberspace operations, is likewise difficult~\cite{Grant23}. The lack of such feedback makes it difficult for a commander to exercise control over an ensuing operation~\cite{KaCo17}. Identifying vulnerabilities in cyberspace systems, especially those that manifest because of the interactions between elements, can also be traced to its extreme variability and inscrutability.  This necessitate the creation of technologies that attenuate the variety of cyberspace, thereby reducing its inscrutability.

A prevailing view among cyberspace experts is that cyberwarfare is a \textit{learning contest}. The malleability of cyberspace means that defensive cyber operations teams rarely encounter cyberspace that conforms to its nominal configuration. Moreover, the malleability means that small changes implemented by such teams can have unintended consequences such as the temporary denial of critical services or unrecoverable data loss. Cyber operators must explore their cyber terrain, and formulate and test hypotheses, given their subject matter expertise and learned familiarity with the terrain itself. Ultimately, the inscrutability of cyberspace limits for operators to undertake comprehensive planning, and expect those plans to be robust.

Anonymous authors from the Netherlands Defence Intelligence and Security Service have asserted~\cite{AnAn22} that the key to cyberspace operations is achieving and maintaining accesses. Whilst the inscrutability of cyberspace creates the opportunity for the stealthy acquisition of accesses, it also accounts for the significant effort required to achieve those accesses. On the other hand, the malleability of cyberspace contributes to the lack of object permanence, which underpins the basis for maneuver warfare~\cite{KaCo17}, and the perishability of advantage in cyberspace operations~\cite{sorensen2010cyber}. Once discovered, accesses can be denied, and exploited vulnerabilities patched. Perishability in turn promotes the need for covertness in cyberspace operations, as much to protect those accesses and exploits as to protect the \textit{modus operandi}~\cite{AnAn22} of the actors. 

However, we challenge some of the assumptions that underpin this access-centric framing. For example, not all accesses or exploits are equally perishable. Operational technology (OT), i.e. those systems that perform useful functions for society other than provisioning of cyberspace such as process controllers in plant equipment, are often patched less frequently. They may have intermittent connectivity with the rest of cyberspace, be difficult to access physically, or run software or firmware that diverges from the conventional installations. As such, these systems may be more or less vulnerable as an attack vector, depending on the objectives of the attacker. Once exploited, it may be cumbersome for the owner to deny the attacker.

Perishability is a principal consideration when threat actors intend to exploit accesses over long periods. Kalloniatis and Bowles~\cite{KaBo24} argue that cyber exploits are largely regarded as scarce and exquisite mission resources. We attribute this scarcity / exquisiteness -- and the concomitant difficulty with cultivating accesses and exploits -- to the inscrutability of cyberspace, and its perishability as a consequence of the malleability of cyberspace. 

Moore~\cite{Moore2019} distinguishes between presence- and event-based military network operations where presence-based operations
\begin{quote}
    \textit{are strategic capabilities that begin with lengthy network intrusions and conclude with an offensive objective}
\end{quote}
and event-based operations
\begin{quote}
    \textit{are directly-activated tactical tools that can be field-deployed by personnel to create localised effects immediately}
\end{quote}
Leveraging Moore's distinction, Kalloniatis and Bowles argue that not all cyberspace operations need to be presence-based. Event-based operations could induce cyber effects to achieve tactical outcomes within a small spatio-temporal context in which the perishability of advantage is not a concern: ephemeral advantage may be sufficient. Moreover, the inscrutability of cyberspace at the tactical edge, whether against OT or information technology, may prohibit both immediate countermeasures and long-term remediation of vulnerabilities.


It is clear that the inscrutability of cyberspace is a limiting factor. The inherent science and technology challenge is how to increase the scrutability of cyberspace. We now turn to issues of agency in cyberspace.

Even assuming that an actor could access all the fundamental bits\footnote{Literally, the ones and zeros.} of cyberspace, which bits should they act upon to achieve advantage? The variety, malleability, velocity and consequential inscrutability of cyberspace means that operators lack the mental models and tools for exercising agency in cyberspace. Variety calculus~\cite{NiCa21} indicates that C2 of cyberspace can only be achieved through mechanisms that attenuate the extreme variety of cyberspace or amplify the variety of the cyber C2 system so as to achieve requisite variety~\cite{Ashby57}. Niven and Capewell contend~\cite{NiCa22} that the extant C2 approach in Western militaries is unsuitable for the modern cyberspace-connected world. We propose that this necessitates new tools and mental models for amplifying the variety of cyber C2 systems and attenuating the variety of cyberspace. Competitive advantage can therefore be gained by creating a cyber C2 construct of requisite variety with cyberspace that overmatches the variety of an adversary's equivalent cyber C2 construct. Variety overmatch promotes inscrutability for the adversary and creates the opportunity for surprise with concomitant outcomes such as loss of initiative and reduced freedom of maneuver~\cite{NiCa21}.

The malleability of cyberspace means that small changes can have dramatic consequences. However, cyberspace is highly contextualised and significant effort is required to tailor operations to the specific use case. Moore~\cite{Moore2019} remarks that
\begin{quote}
\textit{[w]hereas bullets, shells and missiles function as intended against a wide range of possible targets, intangible warfare [of which cyber warfare is one aspect] is unique in such that it may require the development of tools designed to defeat a particular enemy’s specific technology.
}\end{quote}
Threat actors performing presence-based operations require lead times equivalent in the order of many months to years~\cite{AnAn22} to enable reconnaissance, development of tailored technical effects, and to gain access to systems of interest~\cite{Grant23}. Accordingly, presence-based cyberspace operations require long planning cycles and deliberate investment of resources.

The outsized effect of actions within cyberspace can lead to a blurring of the distinctions between the tactical, operational and strategic echelons, which may be referred to as \textit{strategic collapse}~\cite{AnAn22}. The distinctions between these echelons vary from context to context, but typically constitute temporal and geographic bounds on the delegation of authorities: a tactical commander may have authority to operate within specified local terrain for the duration of a specified mission, whereas strategic commanders may be responsible for global matters over strategic time-frames. Cyberspace operations act on cyberspace that typically span jurisdictions with effects that may span time-scales, thereby leading to strategic collapse. Holding all cyber authorities at the strategic level circumvents this collapse, but has the disadvantage of requiring strategic cyber forces to develop requisite variety with all of cyberspace. As this is not possible, a natural consequence of this arrangement is that cyberspace operations will be largely limited to those supporting strategic objectives. 

Besides exacerbating its inscrutability, the velocity of cyberspace necessitates machine speed operations~\cite{sorensen2010cyber}, which mitigates the opportunity for meaningful strategic leadership~\cite{KaCo17}. The C2 literature increasingly refers to non-human intelligent collaborators (NICs)~\cite{KaBo24,TeMe23}, i.e. machine based agents, in order to deal with  the accelerated decision cycles that cyberspace affords~\cite{Routier14}. Adversaries can be expected to use a range of measures based on autonomy or automation \cite{sorensen2010cyber,applegate2012principle}. This implies that autonomous decision-making in particular -- at least at the lowest levels of control -- is imperative for suitably countering adversarial threats and successfully prosecuting mission outcomes. 

Kott \cite{kott2018autonomy,kott2023autonomous} and Scharre et al. \cite{scharre2014robotics} argue that the proliferation of intelligent, autonomous agents is an emerging reality of warfare, and they will form an ever growing fraction of total military assets. The software basis of such agents means that these agents are themselves part of cyberspace, and therefore amplify the variety of cyberspace. Accordingly, the execution of cyber missions may occur, at least in-part, at machine speed, and will require appropriate C2 constructs for planning and execution, with the commensurate mechanisms for managing authorities and permissions. Examplars of advanced peer-to-peer C2 constructs for the control of malicious botnets have been described in the literature~\cite{carvalho2015mira,andriesse2013botnets}, and have been explored in the context of defensive cyberspace operations~\cite{carvalho2016semi,carvalho2013mtc2,kott2023autonomous,uzunov2021aware2}. 

Machine speed C2 may necessitate the realisation of a set of NIC C2 nodes, each of which has a bounded scope within cyberspace. The creation of such C2 nodes amplifies the variety of one's C2 system commensurate with the velocity of cyberspace, however the bounded scope of a given node is intended to ensure that the node has requisite variety within that scope so as to maximise its agency. The corresponding C2 similitude for this set of NIC C2 nodes would be as follows (cf.\cite{kott2014security,stytz2005toward}): 

\begin{itemize}
	\item Command: intent and scope of authority is captured in a machine-interpretable form, such as goals or task structures\footnote{Goals or task structures could be articulated in either imperative or declarative forms, where imperative vs. declarative expressions of intent pertain to automated vs. autonomous cyberspace operations, respectively.}~\cite{lesser2004evolution}, and assigned to C2 nodes.
	\item Control: C2 nodes perform autonomous planning, execution and monitoring of complex multi-step cyberspace operations, consistent with the machine-interpretable codification of intent and scope of authority.
    \item Coordination: the allocation, negotiation and deconfliction of activities within a community of C2 nodes. \item Communication: the machine-machine exchange of information between C2 nodes. 
\end{itemize}

Although the emphasis within these similitudes is machine-centric, the interaction between human commanders / operators and NIC C2 nodes must also be clearly defined with respect to each aspect of command, control, coordination and communication. A necessary and perhaps confronting implication of this -- if we choose to do so -- is the deliberate delegation of authority, responsibility and competency to NICs~\cite{LaSc05}. However, research suggests that there may be a legally robust approach for doing so~\cite{LaLa12}. 

These observations apply not just to cyberspace operations but to all machine-speed operations\footnote{In fact, the execution of any machine-centric control system inherently requires authority to be ceded to the machine. The human supervisor of such a system is both authorising the control system to operate and to exercise authority over those levers over which the control system has agency.}. However, as NICs both constitute cyberspace and perform C2 functions, the deployment and management of NICs may be effectively inseparable from the C2 of cyberspace operations.

A unique property of the conventional approach to cyberspace operations, relative to the other domains of warfare, is the siloing of offensive and defensive operations. There is a perception that such operations are qualitatively different. This perception is well founded: defensive cyberspace operations do not involve the same authorities and -- in the case of presence-based operations~\cite{Moore2019} -- do not involve the significant effort required to gain and maintain access~\cite{AnAn22}. 

Nevertheless, the independent evolution of offensive and defensive cyberspace operations is potentially problematic.
As Grant notes~\cite{Grant23}, offensive cyberspace operations may beget response actions, i.e. offensive cyberspace operations. The open literature suggests that offensive cyberspace operations can focus on disrupting non-cyber capabilities, e.g. rendering financial sectors~\cite{GrKa21} or uranium enrichment centrifuges~\cite{Zetter14} inoperable. When the target is a nation-state with national cyber capabilities, those cyber capabilities have historically been unaffected or can be rapidly reconstituted (cf. malleability and velocity) and therefore able to strike back against the perceived
perpetrator. As an aside, we note that anonymity of actors is another hallmark of inscrutability, and responses to threat actors of uncertain origin is akin to `opening fire on shadows in the fog of war'~\cite{KaCo17}, which leads to disintegration of order, an undesirable amplification of variety and loss of agency.

It follows that defensive cyberspace operations may need to be coordinated with offensive cyberspace operations, e.g. to harden defences in anticipation of a cyber counter-strike. We also note that the artificiality of cyberspace means that operations by either side in a cyberspace conflict have the potential to create and destroy cyberspace, whether own, adversary or in-between. Defensive operations may be essential in order to preserve those parts of cyberspace that are essential to the realisation of offensive cyberspace operations. Conversely, offensive operations may be warranted as a form of active defence. Siloing of defensive and offensive cyber operations, whilst variety reducing, may also reduce agency in cyberspace. 

Finally, we note that while we argue for the need to increase the scrutability and agency for cyberspace operations, we should also characterise the mission of defensive cyber operations as seeking to reduce the scrutability of one's own slice of cyberspace from the perspective of a potential adversary. Similarly, one seeks to minimise the agency of an adversary in one's own cyberspace. The methods for denying reconnaissance, access and lateral movements address these concerns. In an adversarial context, the goal is to retain maximal agency over one's own slice of cyberspace, while denying agency to the adversary.

\section{Final Remarks}
To summarise, cyberspace is an artificial construct that exhibits extreme and potentially unbounded variety, is highly malleable, operates at high velocity, and therefore allows high velocity operations. It is ephemeral and self-referent in that cyberspace can be both created and destroyed by those operations. It serves as a platform for the emergence of new platforms and new behaviours whether \textit{within} cyberspace, regarded as hard cyber, or \textit{through} cyberspace reaching human users, which may be regarded as soft cyber. It follows that cyberspace operations can  influence those behaviours.

Cyberspace underpins the increasing complexity of military operations, resulting in the need for agile C2 structures and processes exhibiting the requisite variety for commanders to maximise their agency in their operating environments. Automated or autonomous C2 processes in the form of NICs provide one means of amplifying variety, but we note that NICs are embedded elements of cyberspace, and therefore vulnerable to cyber attack and worthy of defence.

C2 of cyberspace operations is especially difficult because of the inscrutability of cyberspace, which limits agency therein. We offer an analogy to summarise the nature of cyberspace operations. 

We imagine a network of binary mechanical switches. The network is vast, effectively infinite, forming a fabric of switches. Each switch is connected to a number of other switches, typically a few, but sometimes many\footnote{In graph-theoretic terms, the degree of any given node may follow a power-law distribution.}. Whether a switch is on or off depends on the state of all the other switches to which it is connected. Although this could be a random system, in practice the connectivity is not random: non-trivial patterns of connectivity lead to coordinated behaviors manifesting as dynamic -- sometimes designed-for, sometimes emergent -- patterns of switching. The network exhibits extreme --  potentially unbounded -- variety, high malleability, and its velocity is determined by the speed at which switching occurs.

At any given time, any given human can see a very small subset of the switches, and may be able to change the state of a smaller subset. Human actions in this network amount to finding and flicking a subset of the switches within their control such that their action percolates through the system and, after thousands / millions / billions of switch changes, changes the state of one or more target switches deep within the network.  For example, vulnerability detection requires searching for pathways between the point of access and a targeted cluster of switches that produce an unwanted and un-designed-for change. Such switching pathways are difficult to find\footnote{i.e. computationally expensive.} even if the pathways are visible. Moreover, the interdependent nature of the system means that the actions of others can change the state of intermediate switches and thereby eliminate some pathways for exploitation while creating others. The deliberate elimination of such pathways can be understood as vulnerability patching, which may be easy if the vulnerability pathway can be traced after the fact, but is difficult otherwise.

For a team to have meaningful agency within this fabric of switches they must have requisite variety to understand the structure and behaviour of the system. This affords the ability to forecast how it will evolve as a consequence of the team's action, or the actions of others. The team must also be able to articulate its intentions with respect to desired state of the network of switches, i.e. it must have a clear purpose. Finally, it must have the ability to act upon the network of switches such that its actions induce coordinated, coherent percolation of switch state in fulfilment of that purpose. To do so, it will need control structures of requisite variety to plan and execute those actions.

Whilst framed as an analogy, if we assume that the fabric of switches is logical instead of mechanical then this fabric of switches is literally cyberspace. Each switch corresponds to 1 bit in cyberspace. The interacting and interdependent protocol stacks can be described in terms of the logical bits that describe their state and specified behaviour\footnote{This is the quantitative variety per Ashby's definition~\cite{Ashby57}}. The interactions of the logic with physical electronic devices allows information to be generated, stored and transmitted, and induces both the processing of information and emergent behaviours. Interactions with itself and with human users then allow the creation of more logical structures and behaviours. 

Cyberspace is deeply inscrutable. Nevertheless, cyber practioners have built a wealth of models and tools to increase its scrutability~\cite{JiJaNaGrZaBa22,PaClTrMaWi18,EsPePaCaHiFrSi16,DuBaMo18}. Analytical tools, dashboards, and frameworks such as MITRE ATT\&CK~\cite{StApMiNiPeTh20,mitreattack} and Cyber Kill Chain~\cite{Croom2010} distill the extreme variety of cyberspace into more digestible pieces. In Variety Calculus terms, they are seeking requisite variety through both variety amplification -- more complex tools -- and variety attenuation -- collapsing the complexity of cyberspace into simpler models. However, in spite of these efforts, cyberspace remains more inscrutable than scrutable. Significantly more effort is required.

On the other hand, less effort has been given to maximising agency in cyberspace. The MITRE D3FEND framework~\cite{mitredefend} is a starting point, but control structures and processes are lacking. The challenge is to furnish cyber C2 operators with the requisite variety to act upon cyberspace, exploiting its malleability and velocity for compelling effect. This will require mental models, technologies, structures and processes that achieve not just requisite variety, but requisite malleability and velocity. We aim to address this in future work.

\section*{Acknowledgements}
The authors thank Ian Johnston and Josh Green for sharing insights and feedback on the manuscript.

\bibliographystyle{unsrturl}

\end{document}